\documentclass[12pt]{iopart}

\usepackage{amsfonts}
\usepackage{amssymb}

\usepackage{graphicx}

\newcommand{\defeq}{:=}

\newcommand{\bolds}{\sigma}  % not actually bold

\newcommand{\dos}[1]{g(#1)}

\newcommand{\eqref}[1]{(\ref{#1})}

\newcommand{\text}[1]{\mathrm{#1}}

\newcommand{\metaregion}{\Gamma_{\text{m}}}
\newcommand{\eqm}{\Gamma_{\text{eq}}}

\newcommand{\tmetaregion}{\widetilde{\metaregion}}

\newcommand{\Gammaeq}{\Gamma_{\text{eq}}}
\newcommand{\Gammam}{\Gamma_{\text{m}}}

\newcommand{\Ham}{\mathcal{H}}

\newcommand{\E}[1]{\mathbb{E} \left\{ #1 \right\}}

\newcommand{\figref}[1]{Fig.~\ref{#1}}

\newcommand{\mean}[1]{\left\langle #1 \right\rangle}

\newcommand{\sigmap}{\sigma^\prime}
\newcommand{\sigmaz}{{\sigma_0}}
\newcommand{\ssp}{{\sigma\to\sigmap}}
\newcommand{\sps}{{\sigmap\to\sigma}}
\newcommand{\partder}[2]{\frac{\partial #1}{\partial #2}}
\newcommand{\Prob}{\mathbb{P}}

\begin{document}

%\title{Metastability of Markov processes}
\title{Metastability in Markov processes}

%\author{Hern\'an \surname{Larralde}}
%\author{Fran\c{c}ois \surname{Leyvraz}}
%\author{David P.\ \surname{Sanders}}

\author{H Larralde, F Leyvraz and D P Sanders}
\address{Centro de Ciencias
F\'{\i}sicas, UNAM, Apartado postal 48-3, C.P.\ 62551, Cuernavaca,
Morelos, Mexico}
\eads{\mailto{hernan@fis.unam.mx}, \mailto{leyvraz@fis.unam.mx} and \mailto{dsanders@fis.unam.mx}}
\date{\today}

%\affiliation{\Cuernavaca}

\begin{abstract}
We present a formalism to describe slowly decaying systems in the
context of finite Markov chains obeying detailed balance. We show that
phase space can be partitioned into approximately decoupled regions,
in which one may introduce restricted Markov chains which are close to
the original process but do not leave these regions.  Within this
context, we identify the conditions under which the decaying system
can be considered to be in a metastable state. Furthermore, we show that such
metastable states can be described in thermodynamic terms and define
their free energy.  This is accomplished showing that the probability
distribution describing the metastable state is indeed proportional to
the equilibrium distribution, as is commonly assumed.  We test the
formalism numerically in the case of the two-dimensional kinetic Ising
model, using the Wang--Landau algorithm to show this proportionality
\emph{explicitly}, and confirm that the proportionality constant is as
derived in the theory. Finally, we extend the formalism to situations
in which a system can have several metastable states.
\end{abstract}
 
\maketitle

\section{Introduction}
\label{sec:intro}
The description of macroscopic states has been achieved succesfully in
the case of systems which are in thermodynamic equilibrium. Indeed,
for these, Gibbs' approach via the canonical or other ensembles,
describes in a well-defined manner most equilibrium systems (under
appropriate ergodicity assumptions), and is also well-supported by
experiment. On the other hand, the goal of similarly describing
appropriate macroscopic states for systems out of equilibrium has, so
far, not been achieved in general, and the attempts to do so have met
with serious difficulties.

An interesting intermediate case is found for so-called metastable
systems. These arise typically near first order phase transitions,
when a phase, which is not the thermodynamically most stable one for
the values of the parameters considered, is nevertheless observed to
persist over very long times. A typical instance occurs, for example,
when a system is rapidly quenched from a region in which a given phase
is stable to another region in which the phase is unstable.

Traditionally (see, for example, Maxwell \cite{MaxwellTheoryHeat}), such
states have been considered in a purely thermodynamic
framework. Analytic forms of the free energy, which involved
non-convexities, were taken to reflect physical reality, and the
states showing local thermodynamic stability (for example, those which
have positive susceptibility or compressibility) were identified with
metastable states as observed in nature. This straightforward
explanation, however, suffered a serious blow when it was understood
that short-range systems must necessarily have convex free energies,
as was shown, for example, in \cite{RuelleStatMechBook}. The presence
of non-convex parts in the free energy of a van der Waals system is
hence necessarily linked to the fact that the van der Waals model is
exact only in the limit of long-range potentials.

Nevertheless, a great deal of work followed along these lines. In
particular, Langer \cite{LangerAnnalsPhysics1967} showed how the
original program of Maxwell and van der Waals could be carried out
under quite reasonable assumptions on the nature of the singularities
present near the coexistence curve of two phases. The metastable phase
was there assumed to be destabilized by the possible -- but very
unlikely -- presence of large droplets of the stable phase. That 
droplets of the stable phase are unlikely to arise was explained by
the fact that they must appear previously as smaller droplets, which
are themselves highly unlikely, since they are strongly suppressed by the
effect of surface tension. In such models, the free energy and the
partition function are obtained from the corresponding equilibrium
quantities by a highly non-trivial analytic continuation around an
essential singularity. The reason why the physical metastable free
energy should be identified with the result of such an analytical
continuation, however, is not entirely clear in this approach.
Indeed, this could hardly be otherwise, since the basic objects with
which this framework operates are drawn from equilibrium statistical
mechanics, whereas the metastable state, being subject to decay to a
state very different from itself, is intrinsically a non-equilibrium
object, albeit perhaps a particularly simple one.

In a different line of attack, Penrose and Lebowitz
\cite{PenroseLebowitz1971} studied metastability directly from a
dynamical point of view. Their considerations were limited to rather
particular models, but the principles involved seem capable of great
generalization.  The crucial idea was the introduction of a
\emph{restricted ensemble}, defined in such a way as to prevent
nucleation from ever taking place. If such a dynamics could indeed be
defined in such a way as to have the characteristics of a typical
equilibrium process, then the thermodynamic approach would be, in a
sense, vindicated.

In more recent work, Gaveau and Schulman \cite{GaveauSchulmanJMP1998}
have succeeded in making this quite precise in the very general
framework of arbitrary Markov processes. Their approach consists in
assuming that some non-equilibrium eigenstate of the linear operator
arising in the master equation describing the dynamics has an
anomalously small eigenvalue with respect to all others. In systems
for which the thermodynamic limit has not yet been taken, arguments
taken from the classical theory of nucleation will often strongly
suggest the existence of such slowly decaying states. As to the
thermodynamic limit, it has been forcefully argued in
\cite{GaveauSchulmanJMP1998} that it may in fact be quite misleading
when applied to metastable systems. In their work, they essentially
show the following:

\begin{enumerate}
    \item Under their assumptions, the space of configurations
    separates into two disjoint subsets, which are both almost
    invariant under the dynamics, one of which can be identified as
    the metastable region. In particular, if the system is
    started in this region and allowed to evolve until it
    attempts to leave it, after which it is killed, then the evolution
    of the system is similar to that in which the system evolves in
    the usual manner.

    \item Similarly, they show that the average time to escape from
    the metastable region is very large.
\end{enumerate}

In this work we develop a rather simple formalism, along lines similar
to those of Gaveau and Schulman, that sharpens and extends their
results in various ways; our development was outlined in
\cite{LarraldeLeyvraz2005}.  We confine ourselves to ergodic and
acyclic processes satisfying detailed balance, for which the
stationary distribution can unambiguously be identified with the
equilibirum distribution in the appropriate ensemble of statistical
mechanics.  For such systems, a restricted dynamics in which the
process is reflected, instead of killed, each time it attempts to
leave the metastable region is shown to also be close to the original
process.  (This kind of restriction was also used in a slightly more
specific context in \cite{OlivieriMetastabilityBook}.) This restricted
process reaches an equilibrium state described by a distribution which
is proportional to the equilibrium distribution of the original
unrestricted process. We also show that under suitable conditions, any
initial condition decays quickly into either a metastable state, in
which case the system is described by the equilibrium distribution of
the restricted process, or to equilibrium.

In this respect, another important point in which we sharpen the
results of Gaveau and Schulman is the following: they prove that
\emph{on average} it takes a long time for a system to decay from the
metastable phase to equilibrium. This, however, is not enough to
account for the expected behavior of metastable systems: for example,
if a system were to decay in a time of order one with probability
$1-\epsilon$ and in a time of order $\epsilon^{-2}$ with probability
$\epsilon$, then the average decay time would be of order
$\epsilon^{-1}$, and hence would diverge as $\epsilon \to 0$, yet no
one would call such a system metastable.  Indeed, in agreement with
\cite{OlivieriMetastabilityBook}, one expects metastable states to
reach a (quasi-)stationary regime quickly and then, in a relatively
abrupt maner, decay to equilibrium. The broadness of the distribution
of the time intervals during which the state remains in the metastable
(quasi-)stationary state is what gives rise to the slow decay of the
distribution describing such a state.  Here we show that, under the
assumptions required to define metastability, it is in fact true that
the \emph{probability} of decay in a short time is very small, in
accordance with the intuitive picture given above.
 
These result support the idea behind the restricted process
approach. Furthermore, since we are really dealing with partition
functions of two systems, both of which are equilibrium systems, the
logarithm of the proportionality constant is related to the difference
in free energy between the unrestricted and restricted systems,
corresponding to the stable and the metastable phases respectively.

Another interesting consequence of these observations is the
following: since a metastable system can, to a good approximation, be
described by an equilibrium process over certain time scales and the
usual connections between time correlations and response to small
external perturbations (fluctuation--dissipation theorem) hold exactly
in the restricted dynamics, then, again to a good approximation, they
will also hold in the metastable state. It is therefore legitimate,
say, to measure the frequency dependent susceptibility in a metastable
state by computing the Fourier transform of the magnetization
autocorrelation function \cite{BaezLarralde2003}.

To illustrate some of the results described above, we study a
two-dimensional Ising model subject to an external field. We
parametrize the phase space by reduced variables (in this case
magnetization and total spin--spin interaction energy, which are
adequate for the system sizes we are considering) and evaluate the
equilibrium distribution over the complete parameter space using the
Wang--Landau algorithm.  Within the metastable region, we compare the
equilibrium distribution to the metastable distribution obtained from
Monte Carlo simulations of the kinetic Ising system.  In the
metastable region within the space of reduced variables, we show that
the metastable distribution is indeed proportional to the equilibrium
distribution, with the proportionality constant being as derived in
the theory.

Finally, we extend the formalism to the case in which the system has
several metastable states. This gives rise to minor complications due
to the possibility that the system may decay to equilibrium by passing
through other metastable states.

The outline of the paper is as follows. In Section~\ref{sec:2}, we
review the formalism, which is similar to that used by Gaveau and
Schulman, and the assumptions and notation to be used throughout the
paper. In Section~\ref{sec:3}, we present and derive the results
described above in the case where the system has a single metastable
state. In Section~\ref{sec:4}, we show how our ideas can be applied to
the Ising model, at least for sufficiently small systems. The results
shown in the previous sections can be confirmed using this test
model. In Section~\ref{sec:5}, we discuss the complications appearing
when a finite number of metastable states are taken into account,
instead of only one. Finally, in Section~\ref{sec:6}, we present some
conclusions and outlook.

\section{Theoretical framework}
\label{sec:2}
We set up the description of slowly decaying as well as metastable
states within the general framework of Markov process, which can then
be applied to a large variety of systems.  Given a \emph{finite} set
$\Gamma$ of elements $\sigma$, we consider a continuous time Markov
chain on this set defined by transition rates $W_{\ssp}$.  The
probability $P(\sigma, t)$ to encounter the system at time $t$ in the
configuration $\sigma$ then obeys the master equation
\begin{equation}
\partder{P}{t}(\sigma,t)=\sum_{\sigmap}\left[
W_\sps P(\sigmap,t)-W_\ssp P(\sigma,t)
\right].
\label{eq:2.1}
\end{equation}
If this Markov process satisfies the conditions of ergodicity and
aperiodicity, see \cite{BremaudBook}, which are usually satisfied in
the systems we are interested in \footnote{Note that we are dealing
with finite systems only, so that problems of ergodicity due to, say,
phase transitions do not arise.}, then the probability distribution
$P(\sigma,t)$ approaches a unique equilibrium distribution
$P_0(\sigma)$ as $t\to\infty$.

We will further assume that the system obeys detailed balance. That
is,	
\begin{equation}
W_\sps P_0(\sigmap)=W_\ssp P_0(\sigma)
\label{eq:2.2}
\end{equation}
holds for all $\sigma$.
We rewrite (\ref{eq:2.1}) in the operator form
\begin{equation}
\partder{P}{t}=LP,
\label{eq:2.4}
\end{equation}
where $P$ is a vector with index $\sigma$, and $L$ is a linear
operator on the space of all such vectors.  A scalar product of two
vectors $\phi$ and $\psi$ can be defined as
\begin{equation}
(\phi,\psi) \defeq
\sum_\sigma\frac{\phi(\sigma)\psi(\sigma)}{P_0(\sigma)},
\label{eq:2.5}
\end{equation}
under which, given the detailed balance condition (\ref{eq:2.2}), the
operator $L$ is self-adjoint:
\begin{equation}
(\phi,L\psi)=(L\phi,\psi).
\label{eq:2.41}
\end{equation}

Since the underlying vector space is finite-dimensional, there is a
complete orthonormal set of $N$ eigenvectors $P_n$ satisfying
\begin{equation}
LP_n = -\Omega_n P_n,
\label{eq:2.6}
\end{equation}
where the $\Omega_n$ are by definition arranged in increasing order,
and $N$ is the number of elements of $\Gamma$, i.e.\ the number of possible
configurations. The existence of an
equilibrium distribution implies that $\Omega_0=0$ and the
corresponding $P_0$ is in fact the equilibrium distribution. All other
$\Omega_n$ are strictly positive.

Using the orthonormality of the $P_n$ we can write 
\begin{equation}
(P_0,P_n)=\sum_\sigma P_n(\sigma)=\delta_{n,0},
\label{eq:2.7}
\end{equation}
implying that $P_0(\sigma)$ is normalized and that adding to it
arbitrary multiples of $P_n(\sigma)$, when $n\geq1$, does not alter
this normalization.  The completeness of the eigenvectors
(\ref{eq:2.6}) implies that
\begin{equation}
\delta_{\sigmaz}(\sigma):=\delta_{\sigmaz, \sigma}=\sum_{n=0}^N 
\frac{P_n(\sigma)P_n(\sigmaz)}{P_0(\sigmaz)} .
\label{eq:2.8}
\end{equation}
This leads to an exact expression for the probability
of arriving from $\sigmaz$ to $\sigma$ in time $t$: 
\begin{equation}
P(\sigma,t;\sigmaz,0)=e^{Lt}\delta_{\sigmaz}(\sigma)=
P_0(\sigma)+\sum_{n=1}^N 
\frac{P_n(\sigma)P_n(\sigmaz)}{P_0(\sigmaz)}
e^{-\Omega_n t}.
\label{eq:2.9}
\end{equation}
In the following, we shall say that a system is slowly decaying if at
least one of its eigenvalues $\Omega_{n}$ is much less than all the
others. At first, we shall limit ourselves to the case in which there
is only one slow eigenvalue, that is, when $\Omega_1 \ll\Omega_n$ for
all $n\geq2$.

Now consider a process evolving from the initial condition
$\sigmaz$. Then, from (\ref{eq:2.9}), in the relevant time range
$\Omega_2^{-1}\ll t\ll\Omega_1^{-1}$, one finds that the configuration
$\sigma$ is occupied with the following (time-independent) probability
\begin{equation}
P(\sigma)=P_0(\sigma)+\frac{P_1(\sigmaz)}{P_0(\sigmaz)}P_1(\sigma).
\label{eq:2.10}
\end{equation}
Note that, due to (\ref{eq:2.7}), this is normalized. Also, since it
differs exponentially little from the exact result, we may conclude
that it is positive, except perhaps in some places where it assumes
exponentially small negative values. This situation can be corrected
by setting the negative values to zero and recomputing the
normalization, which leads to negligible alterations.

This result focuses our attention on the value
$P_1(\sigmaz)/P_0(\sigmaz)$, which characterizes the nature of the
initial condition.  This quantity will be central to all that follows.
In particular it will allow us to determine when the initial condition
can be called metastable and the resulting probability distribution
given by (\ref{eq:2.10}) can justifiably be identified with that of a
metastable state. 

In what follows, we denote $P_1(\sigma)/P_0(\sigma)$ by $C(\sigma)$,
and the maximum value of $C(\sigma)$ over all $\sigma \in \Gamma$ by
$C$. Next we define the sets $\Gammam$ and $\Gammaeq$ as follows:
\begin{equation}
\Gammam \defeq \left\{\sigma \in
\Gamma:\frac{C}{2}\leq\frac{P_1(\sigma)}{P_0(\sigma)}
\leq C
\right\},
\label{eq:2.103}
\end{equation}
and $\Gammaeq$ is defined as the complement of $\Gammam$ in
$\Gamma$. We show in Section~\ref{sec:3} that the choice of the factor
$1/2$ to define the lower bound on $C(\sigma)$ in (\ref{eq:2.103}) is
relatively arbitrary.

Equation (\ref{eq:2.103}) defines a partition of phase space into two
disjoint sets. In the following we shall address the question of the
extent to which we can use this partition to define a metastable
state, and in particular, to understand when a standard thermodynamic
approach to the study of such systems is legitimate.

To this end we single out among the slowly decaying systems those
characterized by the condition that the probability of being found
within
$\Gammam$ in equilibrium is negligibly small, i.e.\ such that
\begin{equation}
\mu \defeq \sum_{\sigma\in\Gammam} P_0(\sigma) \ll 1.
\label{eq:2.11}
\end{equation}
We will call metastable systems the slowly decaying systems satisfying
this condition. In particular, from normalization, metastable states
satisfy:
\begin{equation}
\sum_{\sigma\in\Gammaeq}P_0(\sigma) = 1-\mu \approx 1.
\label{eq:2.12}
\end{equation}
We now turn to proving various properties both for slowly decaying
systems not satisfying condition (\ref{eq:2.11}), and for metastable
systems, with a view to justifying the usual assumptions concerning
the description of the latter.

\section{Results and proofs}
\label{sec:3}

We begin by considering a slowly decaying system with a single slow
mode,
so that its phase space is partitioned into $\Gammaeq$ and $\Gammam$
as before. We first consider the case in which the initial condition
$\sigmaz$ satisfies 
\begin{equation}
    C(\sigmaz)=C.
    \label{eq:3.1}
\end{equation}
In this case, we define $Q_{1}(\sigma)$ as the quasi-stationary
distribution which arises from this initial condition over the time
range $\Omega_{2}^{-1}\ll t\ll\Omega_{1}^{-1}$, given by (see
(\ref{eq:2.10}))
\begin{equation}
    Q_{1}(\sigma) \defeq P_{0}(\sigma)+CP_{1}(\sigma).
    \label{eq:3.101}
\end{equation}

\subsection{Probability of exit from the metastable state}
Let us define the random variable $T$ as the time at which a path
starting at $\sigmaz$ satisfying (\ref{eq:3.1}) reaches $\Gammaeq$ for
the first time. A key result is that with high probability this time
is large.  Indeed, for these processes, we have
\begin{equation}
\Prob(T\leq t)\leq2\left( 1-e^{-\Omega_1t} \right)=O(\Omega_1t).
\label{eq:a.8}
\end{equation} 
Here $\Prob(\cdots)$ denotes the probability of an event; for
example, the LHS of (\ref{eq:a.8}) denotes the probability of $T$
being less than $t$.

To prove this, we proceed as follows.  We denote by $\sigma(t)$ the
path followed by the process starting at $\sigma(0) = \sigmaz$ of the
Markov process defined by (\ref{eq:2.1}), and by $\E{\cdots}$ the
expectation value.

The set $\Gammam$ is defined by a condition on the function
$C(\sigma)$, so to study the first exit time $T$ from this set, we
must consider the evolution of $C[\sigma(t)]$ as a function of time.
We thus consider, for $t^\prime > t$,
\begin{eqnarray}
\E{
\left. C[\sigma(t^\prime)] \right| \sigma(t) }
&=\sum\limits_{\sigma^\prime} C(\sigma^\prime)
P(\sigma^\prime, t^\prime|\sigma,t) \nonumber \\ 
&=
\sum\limits_{\sigma^\prime}
\frac{P_1(\sigma^\prime)} {P_0(\sigma^\prime)}\left(P_0(\sigma^\prime)
+\sum_{n=1}^N \frac{P_n(\sigma^\prime)P_n(\sigma)}{P_0(\sigma)}
e^{-\Omega_n (t^\prime-t)}\right)
\label{eq:a.51} \nonumber \\
&= e^{\Omega_1 (t-t^\prime)}C[\sigma(t)],
\end{eqnarray}
where we have used equation (\ref{eq:2.9}) and the definition of
$C(\sigma)$. The last equality follows from the orthonormality
of the basis $P_n(\sigma)$.
Thus we have
\begin{equation}
\E{\left.e^{\Omega_1t^{\prime}}C[\sigma(t^\prime)]\right|\sigma(t)}
=e^{\Omega_1t}C[\sigma(t)],
\label{eq:a.5}
\end{equation}
so that $e^{\Omega_1t}C[\sigma(t)]$ is a martingale
\cite{BremaudBook}; intuitively, this means that, on average, it neither grows
nor decreases with time.

Furthermore, $T$ is a so-called stopping time, that is, it is known at
time $t$ whether the event $T\leq t$ has taken place or not.  If we
now define $\tau=\min(t,T)$, then by standard theorems on martingales
and stopping times (see e.g.\ \cite{BremaudBook}), it follows that
\begin{equation}
\E{ e^{\Omega_1 \tau}C[\sigma(\tau)]}= C(\sigma_0),
\label{eq:a.6}
\end{equation}
where $\sigma_0$ is the initial condition of the process, which we
chose such that $C(\sigma_0) = C$, its maximum possible value.  We can
therefore estimate the LHS of (\ref{eq:a.6}) from above:
%\begin{widetext}
\begin{eqnarray}
\hspace*{-60pt}C=\E{e^{\Omega_1\tau}C[\sigma(\tau)]}&\leq&\frac{C}{2}
\E{\left.
e^{\Omega_1 T}\right| T\leq t }\Prob(T\leq t)+
\E{\left. e^{\Omega_1 t}C[\sigma(t)]\right| T> t }\Prob(T> t)
\nonumber\\
&\leq& C e^{\Omega_1 t}\left\{\frac{1}{2}\Prob(T\leq
t) + \left[1-\Prob(T\leq t)\right] \right\},
\label{eq:a.7}
\end{eqnarray}
from which (\ref{eq:a.8}) follows immediately.
%\end{widetext}

This result is of considerable interest. It represents a significant
sharpening of a result due to Gaveau and Schulman
\cite{GaveauSchulmanJMP1998}, stating that the \emph{average value}
$\langle T\rangle$ is large. Here we show that, for appropriate
initial conditions, the system is very unlikely to leave $\Gammam$
before time $t$ in the relevant time range $t\ll\Omega_1^{-1}$. From
this result one may also derive the following estimate on
$P_{0}(\sigma)$ and $P_{1}(\sigma)$, which will be of use later:
\begin{equation}
\nu \defeq \sum_{\sigma\in\Gammaeq}Q_{1}(\sigma)=
\sum_{\sigma\in\Gammaeq}\left[P_0(\sigma)+CP_1(\sigma)\right]
\leq\Prob(T\leq t) \ll1.
\label{eq:a.33}
\end{equation}
The inequality follows from the fact that $\nu$ is equal to the total
probability of finding a system started at an initial condition
$\sigmaz$ with $C(\sigmaz)=C$ in $\Gammaeq$ at time $t$. But this is
less than $\Prob(T\leq t)$, so that the estimate (\ref{eq:a.33})
follows. If we think of $Q_{1}(\sigma)$ as describing a metastable
state, then (\ref{eq:a.33}) states the (perhaps unsurprising) fact
that the metastable state is entirely concentrated outside $\Gammaeq$.
Note that the converse, namely that $P_{0}$ has only negligible weight
in $\Gammam$ cannot be shown in a similar way. Rather, this condition
is what we introduce in equation (\ref{eq:2.11}) as an
\emph{additional hypothesis} to single out true metastable states from
slowly decaying systems.

\subsection{Definition of restricted process in the metastable state}

We now introduce a restricted Markov process in order to be able to
treat the slowly decaying system as if it were in fact in equilibrium.
To this end, define the following restricted transition rates:
\begin{equation}
W^R_\sps \defeq
\left\{
\begin{array}{ll}
W_\sps&\qquad \sigma,\sigmap\in\Gammam ~~~\mbox{or}~~~
\sigma,\sigmap\in\Gammaeq\cr 0&\qquad\mbox{otherwise}.
\end{array}
\right.
\label{eq:a.21}
\end{equation}
It is clear that the rates $W^R_\sps$ only allow for connections
within $\Gammam$ or $\Gammaeq$. In fact, the $R$ process can be
intuitively understood as a process that imposes reflecting boundary
conditions at the border separating $\Gammam$ from $\Gammaeq$
\footnote{This process differs from the one considered in
\cite{GaveauSchulmanJMP1998}, in which the process is killed whenever
it attempts to leave $\Gammam$.}. Since $P_0(\sigma)$ satisfies
detailed balance in the original process, it is still the equilibrium
distribution for this restricted process, with respect to which it
also satisfies detailed balance. But the restricted system is no
longer ergodic, and therefore $P_0(\sigma)$ is not the unique
stationary distribution. Indeed, $P_1^R(\sigma)$ defined by
\begin{equation}
P_1^R(\sigma) \defeq
\left\{
\begin{array}{ll}
C' P_0(\sigma),&\qquad\sigma\in\Gammam \cr 
S' P_0(\sigma), &\qquad\sigma\in\Gammaeq
\end{array}
\right.
\end{equation}
is stationary for any constants $C'$ and $S'$. In particular,
we may choose these constants so that $\sum_\sigma P_1^R(\sigma)=0$
and $(P_1^R,P_1^R)=1$. This implies that
\begin{equation}
C'=\left(\frac{\sum_{\Gammaeq}P_0(\sigma)}{\sum_{\Gammam}P_0(\sigma)}
\right)^{1/2};\qquad\qquad S'=-1/C'.
\label{eq:a.210}
\end{equation}

Of course, it is now very tempting to identify $P_1^R$ with $P_1$. In
order to do this, we need to show that the process defined by
(\ref{eq:a.21}), which we denote by $R$ (for Restricted), remains
close to the original Markov process defined by the rate $W_\ssp$,
which we denote by $P$ (for Physical).  This can indeed be shown for a
slowly decaying system, if the initial condition $\sigmaz$ satisfies
$C(\sigmaz)=C$ and $t$ is in the relevant time range
$\Omega_{1}t\ll1\ll\Omega_{2}t$.  We define closeness as follows: for
any subset $X\subset\Gammam$, define
\begin{equation}
p_X(t) \defeq \left|\Prob\left\{\sigma_P(t)\in X\right\}-
\Prob\left\{\sigma_R(t)\in X\right\}\right|,
\label{eq:a.22}
\end{equation}
where $\sigma_P(t)$ and $\sigma_R(t)$ are paths of the $P$ and $R$
processes, respectively. We will say that the two processes are
\emph{close in variation} if $p_X(t)$ is small for any
$X\subset\Gammam$.

For the proof, we make the following observation, inspired by the
coupling techniques of probability theory. We define a compound
process $K=(\sigma_P,\sigma_R)$ on the product space
$\Gamma\times\Gamma$ as follows: $\sigma_P$ moves according to the $P$
process, that is, via the rates $W$, and $\sigma_R$ follows $\sigma_P$
around as long as the latter remains in $\Gammam$. As soon as
$\sigma_P$ leaves $\Gammam$, however, each process evolves
independently according to their respective rates.  By construction,
the projections of the compound process $K$ on either subspace yield
the processes $R$ and $P$, respectively.  The two paths $\sigma_R(t)$
and $\sigma_P(t)$ can thus be viewed as projections of the process
$K$.

We wish to show that $\sup\limits_{X\subset \Gamma} p_X(t)$ is small
in the relevant time range. This is achieved as follows: again let us
introduce the random time $T$ as the first time at which
$\sigma_P(t)$, starting from $\sigma_0$ for which $C(\sigma_0)=C$,
leaves $\Gammam$. It then follows by the construction of the process
$K$ that
\begin{eqnarray}
p_X(t) &=&\left|\Prob\left\{\sigma_P(t)\in X | T<t\right\}+
\Prob\left\{\sigma_P(t)\in X | T\geq t\right\}\right.\nonumber\\
&-& \left.\Prob\left\{\sigma_R(t)\in X | T<t\right\}-
\Prob\left\{\sigma_R(t)\in X | T\geq t\right\}\right|\nonumber\\
&=&\left|\Prob\left\{\sigma_P(t)\in X | T<t\right\}-
\Prob\left\{\sigma_R(t)\in X | T<t\right\}\right|\nonumber\\
&=&\left|\Prob\left\{\sigma_P(t)\in X | T<t\right\}-
\Prob\left\{\sigma_R(t)\in X|T<t\right\}\right|\times
\Prob(T<t)\nonumber\\
&\leq&\Prob(T<t),\label{eq:a.23}
\end{eqnarray}
which is indeed small for $t\ll \Omega_1^{-1}$ according to
(\ref{eq:a.8}). As this holds for any $X\subset \Gamma$, the
probability distribution for the restricted process is close in
variation to that of the physical process. 

This implies that within the relevant time range, if
$\sigma_0\in\Gammam$, then
\begin{equation}
P_P(\sigma,t;\sigma_0,0)\approx P_R(\sigma,t;\sigma_0,0),
\label{eq:a.231}
\end{equation}
where $P_{P,R}(\sigma,t;\sigma_0,0)$ denote the transition
probabilities from $\sigma_0$ to $\sigma$ in a time $t$ for the
physical and the restricted process respectively. Again choosing
$\sigma_0$ such that $C(\sigma_0)=C$ and expressing each of these
distributions in terms of the eigenfunctions of their respective
evolution operators, we have:
\begin{equation}
P_P(\sigma,t;\sigma_0,0)=P_0(\sigma)+ CP_1^P(\sigma) e^{-\Omega_1^P t}
+ \sum_{n=2}^N \frac{P^P_n(\sigma)P^P_n(\sigma_0)}{P_0(\sigma_0)}
e^{-\Omega^P_n t}
\end{equation}
and
\begin{equation}
P_R(\sigma,t;\sigmaz,0) = P_0(\sigma) + C'P^R_1(\sigma) e^{-\Omega_1^R
t}+\sum_{n=2}^N \frac{P^R_n(\sigma)P^R_n(\sigma_0)}{P_0(\sigma_0)}
e^{-\Omega^R_n t}.
\end{equation}
Then, the relation expressed in (\ref{eq:a.231}) implies that
$P^R_n(\sigma)\approx P^P_n(\sigma)$, if these quantities are not
negligible in $\Gammam$, in which case we also have
$\Omega^R_n\approx \Omega^P_n$. Thus, again in the time range
$\Omega_2^{-1}\ll t\ll\Omega_1^{-1}$, we are left with
\begin{equation}
Q_{1}^{P}(\sigma) \defeq P_0(\sigma)+CP^P_1(\sigma) \approx
P_0(\sigma) + C'P^R_1(\sigma)=:Q_{1}^{R}(\sigma),
\end{equation}
which, together with the fact that
$\sum_{\Gammam}Q_1^{P}(\sigma)\approx \sum_{\Gammam}Q^R_1(\sigma)=1$,
leads to
\begin{equation}
C\approx C^\prime \qquad {\rm and}\qquad P^P_1(\sigma) \approx
P^R_1(\sigma).
\label{eq:a.232}
\end{equation}
We have therefore two results of interest: on the one hand, the first
passage time from a state $\sigmaz$ satisfying $C(\sigmaz)=C$ to
$\Gammaeq$ is very unlikely to be short. On the other, the process
starting at $\sigmaz$ restricted to remain forever in $\Gammam$ is
quite similar to the original unrestricted process for times in the
relevant time range.

\subsection{Generalisation to other initial conditions}

So far we have restricted attention to initial conditions $\sigma_0$
such that $C(\sigma_0) = C$.  We now show that to a large extent this
requirement on the initial condition becomes unnecessary if one makes
the hypothesis that the system is a metastable one, that is, that
(\ref{eq:2.11}) holds.  For such systems we will prove the following
basic property: no matter what the initial condition $\sigma_0$ is, provided
it satisfies $C(\sigma_0)/C=O(1)$, within a time of order
$\Omega_{2}^{-1}$ the system will either be in $\Gammaeq$ or else it
will satisfy approximately the condition $C[\sigma(t)]=C$. This means,
therefore, that the two results described above can be applied
whatever the initial condition, provided only that the process remains
within $\Gammam$ for a short time.

To show this we first need an auxiliary result, which also depends on
the extra assumption that defines metastability: If we consider an
initial condition $\sigma_0^{(p)}$ such that $C(\sigma_0^{(p)}) =
(1-p)C$, then the probability that this initial condition ends up in
$\Gammaeq$, after a time significantly larger than $\Omega_2^{-1}$ has
elapsed, is $p$. Indeed, in the relevant time range $\Omega_1^{-1}\gg
t\gg \Omega_2^{-1}$, this probability is
\begin{equation}
\hspace*{-60pt}\Prob\left\{\left.\sigma^{(p)}(t)\in\Gammaeq \right|
C[\sigma_0^{(p)}]=(1-p)C\right\}
\approx
(1-p)\sum_{\sigma\in\Gammaeq}
Q_1(\sigma)+p\sum_{\sigma\in\Gammaeq}P_{0}(\sigma)
\approx p,
\label{eq:prob-p}
\end{equation}
where we have combined the facts that for slowly decaying systems
$Q_{1}$ has essentially no weight in $\Gammaeq$ (equation
(\ref{eq:a.33})), and that $P_{0}$ has no weight in $\Gammam$
(equation (\ref{eq:2.11})).

This result further implies that for values of $p$ such that $p/\nu\gg
1$, where $\nu$ is defined by (\ref{eq:a.33}), we have
\begin{equation}
F(p) \defeq \sum_{\sigma : C(\sigma)\leq (1-p)C}
Q_1(\sigma) \ll 1. 
\end{equation}
Indeed, consider a system evolving from an initial state given by
$P(\sigma,0)=Q_1(\sigma)$. The probability of finding the system in
$\Gammaeq$ after a time $t$ will then be given by
\begin{equation}
\sum_{\sigma\in\Gammaeq} P(\sigma,t)=\sum_{\sigma \in \Gamma}
\Prob\left(\sigma(t)\in \Gammaeq|\sigma_0=\sigma\right)P(\sigma,0).
\end{equation}
It then follows that
\begin{equation}
\sum_{\sigma\in\Gammaeq} P(\sigma,t)\geq\sum_{\sigma :
  C(\sigma)\leq (1-p)C} 
\Prob\left(\sigma(t)\in \Gammaeq|\sigma_0=\sigma\right)P(\sigma,0).
\end{equation}
Now, equation (\ref{eq:prob-p}) implies that, in the relevant time
range,
\begin{equation}
\Prob\left(\sigma(t)\in \Gammaeq|\sigma_0=\sigma\right)\geq p, \qquad
\text{if } \quad C(\sigma)\leq (1-p)C,
\end{equation}
so that
\begin{equation}
\sum_{\sigma\in\Gammaeq} P(\sigma,t)\geq p\sum_{\sigma :
 C(\sigma)\leq (1-p)C} P(\sigma,0).
\end{equation}
However, since the initial state was $Q_1$, which is essentially
stationary in this time range, we have $P(\sigma,t)\approx
Q_1(\sigma)$, giving
\begin{equation}
p F(p) \leq \sum_{\sigma\in\Gammaeq} Q_1(\sigma) = \nu.
\end{equation}
Since $\nu$ is negligibly small, we thus find that $F(p) \leq \nu/p
\approx 0$.

In a similar way, we can show that $P_0(\sigma)$ is non-negligible
only for the states for which $C(\sigma)\approx 0$. This time consider
\begin{equation}
G(p) \defeq \sum\limits_{\sigma : (1-p)C\leq C(\sigma)}P_0(\sigma).
\label{eq:a.150}
\end{equation}
After a time of order $\Omega_2^{-1}$, at least $(1-p)G(p)$ of
these states will end up in $\Gammam$. Thus, following the same
line of reasoning as before, we can conclude that
\begin{equation}
(1-p)G(p) \leq \sum\limits_{\sigma\in\Gammam}P_0(\sigma) = \mu .
\end{equation}
But our basic hypothesis is that for metastable states, $\mu$ is
negligibly small, thus $G(p)\ll 1$ for $p\ll 1-\mu$. In other words,
if the condition for metastability, equation (\ref{eq:2.12}), holds
when $\Gammam$ is defined by the inequalities (\ref{eq:2.103}), then a
similar claim can be shown when the prefactor $1/2$ is replaced by
essentially any other number well within $\nu$ and $ 1-\mu$. This
means, in fact, that outside a boundary set with relatively small
measure both with respect to $P_{0}$ and to $Q_{1}$, the function
$C(\sigma)$ takes only the values $0$ and $C$.

Conversely, it is obvious that if $G(p)\ll 1$ for all $p$ within $\nu$
and $1-\mu$, then the state will be metastable in the sense that
equation (\ref{eq:2.12}) is satisfied.  No similar converse statement
holds for $F(p)$: in that case, $F(p)$ was found to be negligible as a
consequence of the fact that the probability of leaving $\Gammam$
within the relevant time range is negligibly small. This, as we have
seen, is the case for arbitrary slowly decaying systems, if the
initial condition $\sigmaz$ satisfies $C(\sigmaz)=C$. Thus, for
non-metastable slowly decaying states, $F(p)$ would be negligible due
to the very slowness of their decay. However, such modes would not
correspond to metastable states unless assumption (\ref{eq:2.12})
held, and thus, $G(p)\ll 1$.

For non-metastable slowly decaying systems, we have that $G(p)$ is not
negligible, indicating that one can find states $\sigma$ in
equilibrium with any value of $C(\sigma)$, including $C(\sigma)\approx
C$, all of which would decay slowly.  For such initial conditions, the
almost certain absence of decay within the relevant time range cannot
be expected. In fact, instead of reaching a stationary state which
suddenly decays, the properties of these systems will evolve
continuously in time until they reach equilibrium.  Physical examples
of such slowly decaying systems are hard to come by: the obvious
instances that come to mind (slow hydrodynamic modes, such as
diffusion, necessary to reach a uniform equilibrium from a long-
wavelength perturbation) almost invariably involve a quasi-continuum
gapless spectrum near zero, and are thus ruled out by our basic
assumption. On the other hand, a trivial, though unenlightening,
example shows that non-metastable but slowly decaying states do exist:
if one specific spin in an Ising model is flipped at a much slower
rate than all others, it will, as is easily verified, create a slowly
decaying eigenstate which is not metastable in the sense that it does
not satisfy (\ref{eq:2.11})

\subsection{Structure of metastable states}

The picture that emerges then, is that for systems having metastable
states, after a relatively short transient time, the system will only
be found  in states $\sigma$ for which either $C(\sigma)\approx 0$
(equilibrium) or $C(\sigma)\approx C$ (metastable), independently
of the initial condition. Further, the dynamical behavior is
described to a good approximation by the restricted Markov process
which is reflected whenever it attempts to go from $\Gammam$ to
$\Gammaeq$.

Finally this can be interpreted as follows: the state in which a
metastable state remains throughout the relevant time range
$\Omega_{1}t\ll1\ll\Omega_{2}t$ is determined by $Q_{1}$, which is in
principle defined in a way that depends on the dynamics. However, as
we have seen, it turns out that
\begin{equation}
    P_{1}(\sigma)=CP_{0}(\sigma)\qquad \text{for} \quad
\sigma\in\Gammam,
    \label{eq:3.20}
\end{equation}
from which immediately follows
\begin{equation}
    Q_{1}(\sigma)=Z_{1}P_{0}(\sigma)\qquad \text{for} \quad
\sigma\in\Gammam,
    \label{eq:3.201}
\end{equation}
where $Z_{1}=1+C^2$.

It follows that the only influence of dynamics on the metastable state
is that it defines the extent of $\Gammam$. In other words, the
equilibrium ensemble restricted to a suitable subset $\Gammam$ of
phase space describes the metastable state. From this follows, in
particular, that one may straightforwardly define thermodynamic
quantities such as the partition function by
\begin{equation}
    Z_{\text{m}}\defeq\left(\sum_{\sigma\in\Gammam}P_{0}(\sigma)
    \right)^{-1}=Z_{1},
    \label{eq:3.21}
\end{equation}
where the last equality follows from the normalization of $Q_{1}$ and
equation (\ref{eq:3.201}), as well as the fact that, as follows from
(\ref{eq:2.11}), the term $\sum_{\sigma\in\Gammam}P_{0}(\sigma)$ is in
fact negligible. Note that this implies in particular that $C\gg1$.

Similarly, we can show that the fluctuation--dissipation theorem, see
for example \cite{BaezLarralde2003}, will hold for metastable states,
since the dynamical correlation functions over the relevant time range
will be described by a Markov process close to the one that reflects
the system back to $\Gammam$ whenever it attempts to leave it.  This
process, however, is a well-defined Markov process satisfying detailed
balance which has the normalized restriction of $P_{0}(\sigma)$ to
$\Gammam$ as an equilibrium state, so that the
fluctuation--dissipation theorem can be shown for it in a
straightforward way.

\section{An illustration: The kinetic Ising model}
\label{sec:4}

We now proceed to show how these ideas can be applied concretely in
the case of the two-dimensional Ising model.  Here, the
configurations are collections $\bolds=(\sigma_i)_{i=1,\ldots,N}$ of
spins $\sigma_i = \pm 1$ at site $i$, with energy given by the
Hamiltonian
\begin{equation} \label{eq:hamiltonian}
\Ham(\bolds) = -\sum_{\langle i,j \rangle} \sigma_i \sigma_j - h \sum_i \sigma_i =:
E(\bolds) -h M(\bolds),
\end{equation}
where the first sum is over nearest neighbours in an $N \defeq L
\times L$ square lattice with periodic boundary conditions, and $h$ is
the external magnetic field. $E(\bolds)$ and $M(\bolds)$ are, respectively,
the spin--spin interaction energy and the magnetization of the
configuration $\bolds$.

To obtain a kinetic model which can exhibit metastability, we must
impose a dynamics on the system. For concreteness we use discrete-time
Metropolis spin-flip dynamics \cite{NewmanBarkemaBook}: spin flips are proposed
at random, and
accepted with probability $\min\{1, \exp(-\beta \Delta H)\}$, where
$\Delta H$ is the change in the Hamiltonian \eqref{eq:hamiltonian} due
to the flip, and $\beta \defeq 1/T$ is the inverse temperature
\cite{NewmanBarkemaBook}. Note, however, that the only thing expected
to change under a different local \footnote{This caveat is necessary
since certain non-local dynamics for the Ising model, such as the
Swendsen--Wang algorithm \cite{NewmanBarkemaBook}, suppress metastability
altogether.}
dynamic rule is the extent of the metastable region.  The Metropolis
dynamics gives a discrete-time Markov chain with a unique equilibrium
distribution at fixed $T$ given by the canonical distribution
\begin{equation}\label{eq:gibbs-distribution}
P_0(\bolds) = \frac{1}{Z} \exp[ -\beta \Ham(\bolds) ],
\end{equation}
where 
\begin{equation}
Z \defeq \sum_\bolds \exp[ -\beta \Ham(\bolds)]
\end{equation} 
is the partition function, from which we can obtain all thermodynamic
information at equilibrium.  The Hamiltonian \eqref{eq:hamiltonian},
together with such a spin-flip dynamics, gives the kinetic (or
stochastic) Ising model.

As is well known, if we fix a subcritical temperature $T<T_{c}$, and a
weak external magnetic field $h$ is applied, taken negative
(downwards) without loss of generality, then the spontaneous
magnetization in equilibrium points in the direction of that
field. However, if we initialize the system with all spins up, then
for a broad range of parameters, the system remains in this
thermodynamically unfavorable positively magnetised state for a given
(random) length of time, whose mean depends on the temperature $T$ and
the external field $h$ \cite{RikvoldLifetimes1994}. This state is the
prototype of the metastable states we aim
to describe.

Since the Metropolis Markov chain is ergodic and acyclic
\cite{BremaudBook}, the formalism developed in the previous sections
(when rewritten for discrete-time systems) applies to this system.
Intuitively, it is clear that the kinetic Ising model started in the
metastable region has a hierarchy of relaxation times, with one (the
escape time from the metastable region) being much longer than the
others. Assuming that this is reflected in the spectral properties
required in the derivations above, in this section we show that the
formalism indeed provides a good description of this metastable state.
We remark that many rigorous results have been proved on metastability
in the Ising model in the low-temperature limit: see
\cite{OlivieriMetastabilityBook} for a comprehensive review; in
particular, the separation of eigenvalues required in our formalism
has been proved in this limit in
\cite{BovierCMP2002,BovierGlauberLowTempJSP2002}.  However, our
formalism is valid for any temperature, provided that the eigenvalue
separation is satisfied.

\subsection{Reduced phase space}

To obtain confirmation of the results of Section~\ref{sec:3} in the
case of the kinetic Ising model, we must identify the metastable and
equilibrium regions $\Gammam$ and $\Gammaeq$ and compare the
equilibrium and metastable distributions in each of these
regions. However, given the huge size of phase space even for small
systems, this program cannot be carried out. Instead we must resort to
a reduced description of the complete phase space in terms of a few
variables which, if accurate enough, will reflect the relations we
predict over the complete phase space.

Due to the numerical techniques we use (discussed below), we are
restricted to studying relatively small systems.  For ferromagnetic
Ising models of such sizes, it follows from elementary nucleation
theory that $E$ and $M$ are sufficient to characterize
$\Gammam$. Indeed, we know that nucleation occurs whenever a droplet
of approximate size $R_{c}(\beta, h)$ arises spontaneously, where
$R_{c}$ depends on $\beta$ and $h$, but not on the size of the
system. Inside the critical droplet, the magnetisation has
approximately its equilibrium value, whereas outside it has the
(generally quite different) metastable value. For small systems, it is
therefore generally impossible for a critical droplet to appear
without significantly modifying the magnetisation $M$. For larger
systems, it would be necessary to restrict not only $M$, but also all
magnetisations restricted to cells of size of order $R_{c}$; such
restrictions presumably define $\Gammam$. This has been treated in
detail in particular in \cite{PenroseLebowitz1971}. In the following,
since we are limited to small systems, we decribe $\Gammam$ entirely
in terms of $E$ and $M$.

We refer to the set $\{\bolds \in \Gamma: E(\bolds)=E; \, M(\bolds)=M \}$ of
configurations with given values of the macroscopic variables $E$ and
$M$ as the $(E,M)$ \emph{macrostate}. In this section we work
exclusively on a coarse-grained level in terms of such macrostates,
for the reasons just described, by summing over all configurations
$\bolds$ belonging to a macrostate.  For example, summing
\eqref{eq:gibbs-distribution} over the $(E,M)$ macrostate, we obtain
\begin{equation}
    P_{0}(E,M)=\frac{\dos{E, M}
    \exp\left[ -\beta( E- hM) \right]}{Z(\beta, h)},
\end{equation}
where 
\begin{equation}
\dos{E,M} \defeq \sum_{\bolds\in\Gamma} \delta[E(\bolds)-E] \, \delta[M(\bolds)-M]
\end{equation}
 is the degeneracy (`density of states') of the macrostate $(E,M)$,
i.e.\ the number of configurations $\bolds$ with energy $E$ and
magnetisation $M$, and the partition function can be written as
$Z(\beta, h) = \sum_{E} \sum_{M} \dos{E,M} \exp[-\beta(E-hM)]$.  This
approach was previously used in the context of metastability in
\cite{ShtetoPRE1997}; see also
\cite{SchulmanCoarseGrainsFoundPhys2001} for a method to derive
suitable coarse-grained quantities.

This is a useful representation, since we can compute the joint
density of states $\dos{E,M}$ numerically using the Wang--Landau
algorithm \cite{WangLandauPRL2001, WangLandauPRE2001}; we use a more
efficient version of this algorithm given in
\cite{ZhouBhattUndersandingWL_PRE2005}.  (Computing $\dos{E,M}$
analytically would be equivalent to solving the Ising model in
external field, a still-unsolved problem.)  The fact that we require
the joint density of states as a function of the two parameters $E$
and $M$ restricts us to small systems
\cite{ZhouLandauJointWL_PRL2006}, but we can obtain $\dos{E,M}$
relatively easily for a system of size $32\times 32$ spins, where
metastability can be clearly seen under Metropolis dynamics.  All
numerical results we present are for this system size, for which the
range of possible values for $E$ is $[-2048,2048]$, and for $M$ is
 $[-1024,1024]$. Simulation
times are measured in Monte Carlo steps per spin (MCSS). 

\begin{figure} 
\includegraphics[scale=1]{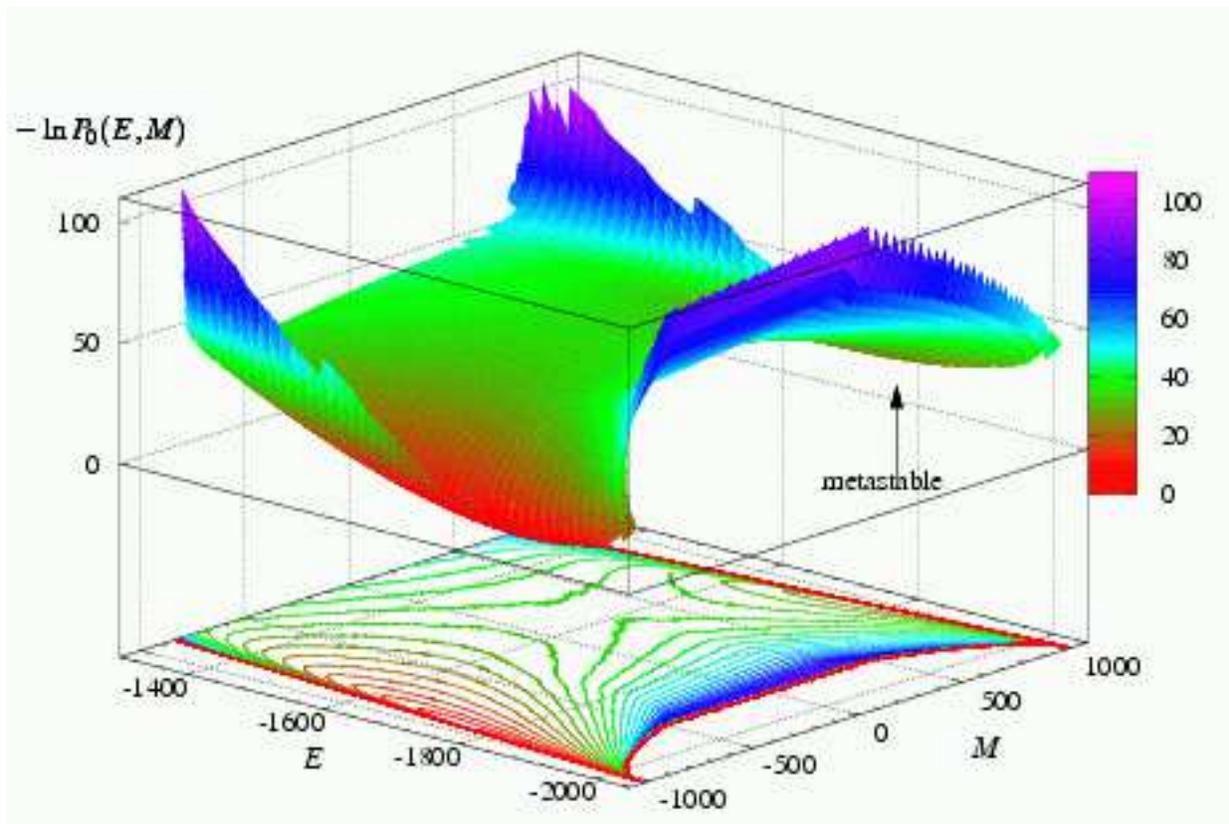}
\caption{\label{fig:eqm-distn} Part of the `free energy' $-\ln
P_0(E,M)$ as a function of $E$ and $M$ for $\beta = 0.5$ and
$h=-0.02$, evaluated from $g(E,M)$ data obtained using the
$2$-parameter Wang--Landau algorithm.  Two minima can be seen: the
higher, metastable minimum is marked. Note
that the $z$ axis
is logarithmic, so that there is a difference of  many orders of magnitude
between their heights. A contour plot is also shown; here the
saddle point, the two minima, and the non-existence of certain $(E,M)$
macrostates are visible. }
\end{figure}

From $\dos{E,M}$ we can obtain the complete partition function, and
hence all thermodynamic information at equilibrium  for given values
of $\beta$ and $h$ \cite{WangLandauPRE2001}.  For example,
\figref{fig:eqm-distn} shows the shape of the equilibrium distribution
$-\ln P_0(E, M) = -\ln g(E,M) + \beta(E-hM) + \ln Z$ for a particular
$\beta$ and $h$ for which a metastable state exists.  

Two minima of different heights can be seen, separated by a saddle;
the higher minimum corresponds to the metastable state, and the lower
one to the equilibrium state. \figref{fig:eqm-distn} can be viewed
as a `free energy' landscape.  If the system starts in the metastable
state, then in order to escape to equilibrium, it must pass over the
free energy barrier near the saddle point \cite{ShtetoPRE1997,
Schulman1980}.

\begin{figure} 
\includegraphics[scale=1]{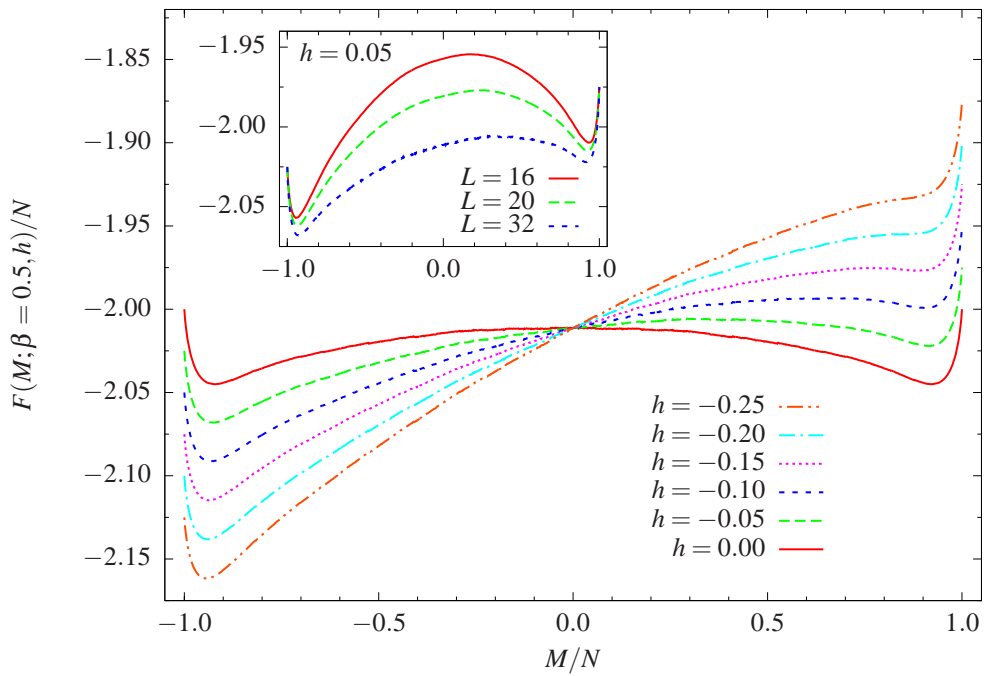}
\caption{\label{fig:F_of_M} Free energy per spin $F(M; \beta, h)/N$ as a
function of magnetisation per site $M/N$ for $L=32$, $\beta = 0.5$ and several
values of $h$, obtained using the
$2$-parameter Wang--Landau algorithm. Again a metastable and an equilibrium
minimum are visible; the former disappears for sufficiently large $|h|$.
The inset is a comparison of $F(M; \beta=0.5, h=0.05)/N$ for different system sizes$L=16, 20, 32$  again as a function of $M/N$, showing the convergence towards a convex function as the
thermodynamic limit is approached ($L \to \infty$).
}
\end{figure}

We remark that an alternative coarse-graining has also been
used  to study metastability in the Ising model, using only
the magnetisation as a coarse-grained quantity.  This can be motivated
by considering the
Ising model in the lattice gas representation, that is, with spin $1$ 
representing a particle and spin $-1$ a void. In that case, the
canonical ensemble is one in which $M$ and $\beta$ are
constant, and the
free energy is given by
\begin{equation}
    F(M; \beta)=-\frac{1}{\beta}\ln\sum_{E}g(E, M) \e^{-\beta E}.
    \label{eq:new1}
\end{equation}
Returning to the Ising model, if we now impose a magnetic field $h$, then the corresponding free energy becomes
$F(M; \beta, h) = F(M; \beta) - hM$. 
This can be obtained from the distribution of Figure \ref{fig:eqm-distn} by summing over all $E$; it is plotted in
\figref{fig:F_of_M}.  $F(M)$ is proportional to the logarithm of the
distribution of the order parameter $M$ \cite{Schulman1980,
TomitaRelaxationMetastableIsing_PRB1992} and can be calculated using
several Monte Carlo methods \cite{ChandlerBook,
BustillosReconstructingFreeEnergy2004}.  The 
Wang--Landau method again has the advantage that we can calculate $F(M; \beta,h)$ for
any parameters $\beta$ and $h$, from a single run.

We see that
the free energy $F(M)$ is still significantly non-convex. This does not, of
course, contradict the rigorous results of \cite{RuelleStatMechBook}, which show that the free energy per 
spin must be convex  in the thermodynamic limit. Indeed, the inset of \figref{fig:F_of_M} illustrates how
this convexity is approached as the system size $L$ increases.  However, it shows
that our simplified description cannot hold for arbitrarily large
systems. As  mentioned previously, to describe the metastable region
adequately, we need to use macrostates specific enough to decide
whether a critical droplet is present or not. What we are suggesting,
however, is that a finite system described in this fashion will
display significant non-convexities in the free energy as defined
here, since there will always be a local minimum corresponding to the
metastable state $Q_{1}(\sigma)$.

\subsection{Finding the metastable region and calculating $C(\sigma)$}

We are interested in the structure of metastable states.  According to
our formalism, such states, when they exist, should be described by
$Q_1(\sigma)=Z_1 P_{0}(\sigma)$ for configurations $\sigma$ in the
metastable region $\Gammam$, with $C$, and hence $Z_1$, being
\emph{constant} over this region.  To test this, we again look at the
coarse-grained level, summing over the $(E,M)$ macrostate to give
\begin{equation}
Q_1(E,M) = Z_1(E,M) P_0(E,M) = [1 + C(E,M)^2] P_0(E,M),
\end{equation}
where $C(E,M)$ and $Z_1(E,M)$ are the mean values of $C(\sigma)$ and
$Z_1(\sigma)$, respectively, for $\sigma$ in the $(E,M)$ macrostate,
and $Q_1(E,M)$ is the sum of $Q_1(\sigma)$ for $\sigma$ in that
macrostate.  Taking logarithms and using $\ln [1 + C(E,M)^2] \simeq 2
\ln C(E,M)$ for $C(E,M)$ large, we obtain
\begin{equation} \label{eq:C_of_E_and_M}
\ln C(E,M) = \frac{1}{2} \left[ \ln Q_1(E,M) - \ln \dos{E,M} +
\ln Z + \beta(E-hM) \right].
\end{equation}

If the theory is correct and, furthermore, if the parameters $E$ and
$M$ provide an adequate representation of the complete phase space of
the system we are studying, then for an $(E,M)$ macrostate whose
configurations $\sigma$ are all in the metastable region $\Gammam$, we
expect that $C(\sigma)=C$ is constant over the macrostate, so that
$C(E,M) = C$.  We thus expect to have a large region in a plot of
$C(E,M)$ where it is essentially constant, i.e.\ a plateau.  This
region of $(E,M)$ space, which we denote by $\tmetaregion$, then
corresponds to the metastable region $\Gammam$ in the complete phase
space.

\begin{figure}
\includegraphics[scale=1]{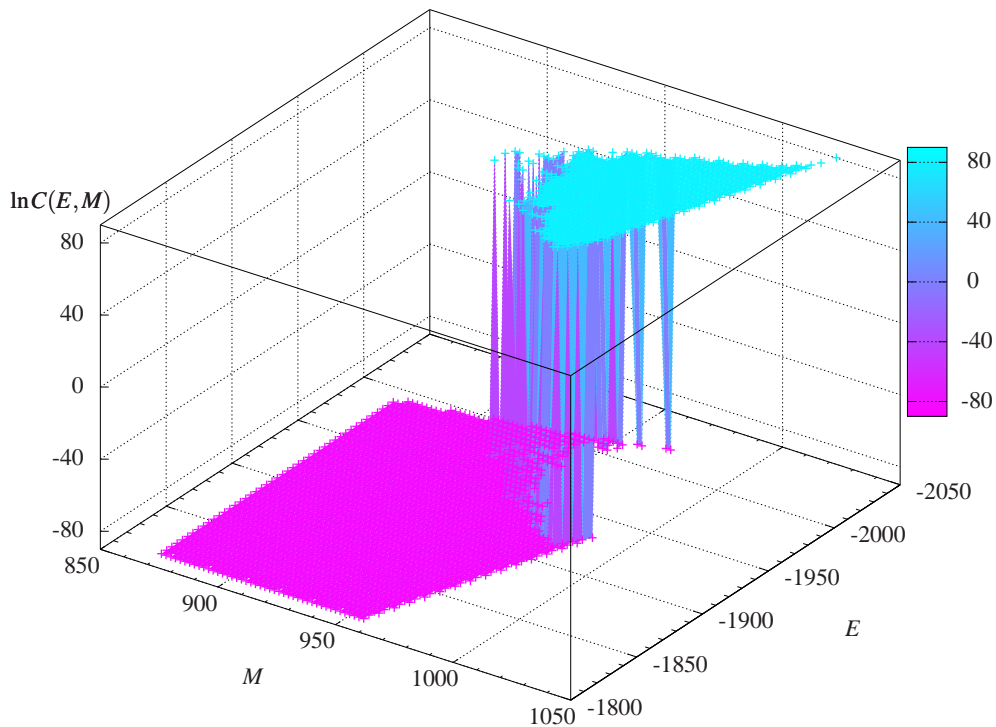}
\caption{\label{fig:C}$\ln |C(E,M)|$ as a function of $E$ and $M$
near the metastable region for $\beta=0.8$ (i.e.\ $T \simeq 0.55
T_c$) and $h=-0.1$, calculated using \eqref{eq:C_of_E_and_M}, from a
single run of $7 \times 10^8$ MCSS with no nucleation events. Outside
the metastable region, $-\ln C$ is shown for comparison.}
\end{figure}

To find this metastable region $\tmetaregion$ in the reduced parameter
space (for given values of $\beta$ and $h$), we must obtain the
metastable probability distribution $Q_1(E, M)$, i.e.\ the probability
that the system is in the $(E,M)$ macrostate while it remains in the
metastable state.  To do so, we record a histogram of the number of
visits to each $(E,M)$ pair while the system remains in the metastable
state, averaging over different runs if necessary.  Normalising this
histogram then gives an estimate of the probability distribution
$Q_1(E,M)$. It is very strongly peaked in a small region of the
$(E,M)$ plane: for example, for the parameters used in \figref{fig:C},
the maximum value of $Q_1$ occurs at $(E_0,M_0) = (-2040, 1022)$, and
is given by $Q_1(E_0, M_0) = 0.218$, so that the system is in this
single macrostate for nearly a quarter of the time spent in the
metastable state; and adding another two macrostates gives more than half
the total probability. Intuitively, the metastable region
$\tmetaregion$ should consist of those $(E,M)$ pairs which have an
appreciable $Q_1$ probability.

We now use the Wang--Landau algorithm to calculate the joint density
of states $\dos{E,M}$ and the partition function $Z$ for the same
lattice for which we calculated $Q_1(E,M)$, and substitute these
values into \eqref{eq:C_of_E_and_M} to obtain $\ln C(E,M)$ as a
function of $E$ and $M$.  Note that  this key application of the
Wang--Landau algorithm  determines $E$ and $M$ as the
macroscopic variables to be used.

\figref{fig:C} shows a plot of $\ln |C(E,M)|$ for values of $\beta$ and $h$ such
that no nucleation event occurred during the (long) simulation, so that the
system was always in the metastable state.  In confirmation of the
theory, a large plateau is apparent.  For some $(E,M)$ macrostates,
$C(E,M)$ is larger than this plateau value. This happens, even though
according to the theory it cannot since the plateau value of $C$ is
its largest possible value, due to the fact that these macrostates are
visited very rarely during the simulation, so that good statistics
cannot be acquired, and their measured frequency is larger than their
true frequency.  Outside the metastable region accessible in the
simulation, we plot $-\ln C$ for comparison, since there we expect
that $Q_1(E,M) = -1/C$ (see \eqref{eq:a.210}).  This neglects the
boundary region between the two phases, to which we have no access
using this method. In the next subsection we present an alternative
approach.

Furthermore, the plateau value of $C$ should be expressible in terms
of the equilibrium distribution $P_0$ as (c.f.\ \eqref{eq:a.210})
\begin{equation}
\label{eq:C-as-free-energy-diff}
\ln C = \frac{1}{2} \left[ \sum_{(E,M) \notin \tmetaregion} \ln
P_0(E,M) - \sum_{(E,M) \in \tmetaregion} \ln P_0(E,M) \right],
\end{equation}
which can be interpreted as the difference in free energy between the
equilibrium and metastable phases.  Indeed, for the case shown in
\figref{fig:C}, the plateau value calculated from the metastable
distribution $Q_1$ using \eqref{eq:C_of_E_and_M} at $(E_0, M_0)$
(where the statistics are best) is $\ln C(E_0,M_0) = 81.595$, whereas
the free energy difference \eqref{eq:C-as-free-energy-diff} gives $\ln
C=81.591$.  Note that if we write $\ln C$ as a free energy difference,
then it is entirely determined by the equilibrium distribution.  The
effect of the dynamics is hidden in the determination of the
metastable region $\tmetaregion$.

We remark that for higher
temperatures, the system does escape from the metastable state during
a run.  In this case, the identification of the metastable region
$\tmetaregion$ is less obvious.  We take it as being those $(E,M)$
with $C(E,M)$ within $\pm 1$ of $C(E_0, M_0)$.

These results provide numerical confirmation that the metastable
distribution $Q_1$ is proportional to the equilibrium distribution
$P_0$ in the metastable region, and that the proportionality constant
$C$ can be related to the difference in free energy between the two
phases.

\subsection{Structure of $C(\sigma)$}

To gain more insight into the function $C(\sigma)$, we can use the
result \eqref{eq:prob-p}, which shows that if we start from an
initial configuration $\sigma_0$ such that $C(\sigma_0) = pC$, then
the probability that after a short relaxation time the system is in
the metastable state is $p$, while the probability that it is in
equilibrium is $1-p$.  We cannot calculate values of $C(\sigma)$
directly, other than in the metastable region $\tmetaregion$, but we
can use this result `in reverse' to obtain a coarse-grained picture
of $C(\sigma)$, as follows.

For each $(E,M)$ macrostate, we wish to generate configurations
$\sigma$ lying in that macrostate.  This is non-trivial, but can be
accomplished by starting from a random initial configuration, with
each spin being up or down with probability $1/2$.  From there we
propose random spin flips, accepting only those which move us towards
the desired value of $(E,M)$.  This process may get stuck, however,
before reaching $(E,M)$, in which case we employ Wang--Landau sampling
(which is known to explore parameter space reasonably efficiently
\cite{WangLandauPRE2001}) in a window containing the current and
desired $(E,M)$ values, to force the system into a configuration
belonging to the required macrostate.  If this does not succeed after
a certain number of steps (for example if there are no configurations
in the target `macrostate'), then we continue with the next
macrostate.  We cannot guarantee that this procedure samples initial
configurations within $(E,M)$ uniformly, but empirically this seems to
be the case, with no particular bias in the
procedure.

We start with $n_0$ configurations within the $(E,M)$ macrostate as above,
evolve each under Metropolis dynamics for a short time, and record
whether the system has reached the equilibrium state, taken to be
configurations with $M(\sigma)\le0$, or not.  The ratio
$n_{\textrm{eq}} / n_0$ of the number of times equilibrium is reached
to the total number of trials is an approximation to $1-p(E,M)$ for
that macrostate.  Note, however, that the fact that we average over
macrostates means that we may not correctly identify the boundaries of
the metastable region: a single macrostate may contain some
configurations which always lead to the metastable state, and others
which always lead to equilibrium, for example.  Nonetheless it gives a
clear picture of the metastable and equilibrium regions, and an idea
of the structure of the boundary between them.

\begin{figure}
\includegraphics[scale=1]{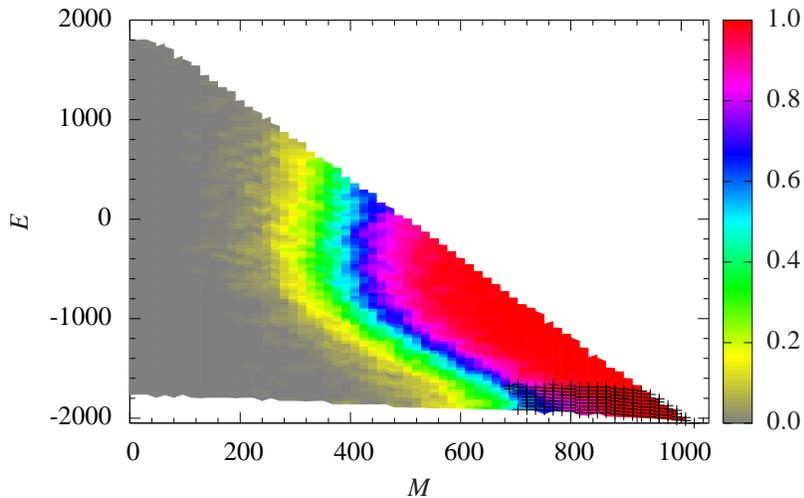}
\caption{\label{fig:landing-prob} Probability $p(E,M)$ of reaching
the metastable state starting from the $(E,M)$ macrostate, for
$\beta=0.6$, $h=-0.1$, and $n_0 = 50$ trials for each $(E,M)$. The crosses at
 bottom right indicate the extent of the metastable region obtained 
from Monte Carlo
simulations, defined as those $(E,M)$ having $\ln C(E,M)$ within $\pm \alpha$ of
$\ln C(E_0, M_0)$, with $\alpha=1$. Changing the tolerance $\alpha$ in the
definition changes the horizontal extent of this region.}
\end{figure}

\figref{fig:landing-prob} shows $p(E,M)$ calculated in this way. We
see clear metastable ($p=1$) and equilibrium ($p=0$) regions,
separated by a boundary region where $p$ takes intermediate values.
The boundary region is larger than we might expect, due to the
smoothing described above, but the system spends little time in this
transition region when the dynamics is taken into account. Note,
however, that according to the results in Section~\ref{sec:3},
exactly where we impose the boundary between the metastable and
equilibrium regions does not affect the results.

Also shown in the figure is the metastable region obtained in Monte Carlo
simulations, as described in the previous subsection.  Note that the
region of $(E,M)$ with $p(E,M) \lesssim 1$ is significantly larger
than this latter definition of the metastable region.
This reflects the fact that there are configurations $\sigma$ which
are never reached from an initial configuration with all
spins up, since the probability of doing so is negligible, and yet
which will decay into the metastable state if started there, thus
belonging to the metastable region according to our definition.  It should also
be noted that the boundary of the region from the simulations lies at
approximately $p(E,M) = 0.5$, and does not significantly vary if
the exact definition of the region is changed, in accordance with the results
of Section~\ref{sec:3}.

\subsection{Relation of $C$ to hysteresis loops}

Since $\ln C$ corresponds to a difference in free energies,
differentiating it with respect to the external field $h$ gives a
difference in magnetisations between the two regions:
\begin{equation}
\frac{\partial(\ln C)}{\partial h} = \frac{\beta}{2}\left(
\mean{M}_{\text{eq}} - \mean{M}_{\text{m}} \right),
\end{equation}
where $\mean{M}_{\text{eq}}$ denotes the mean magnetisation in
equilibrium, given by
\begin{equation}
\mean{M}_{\text{eq}} \defeq \frac{\sum_{\sigma \in \eqm} M(\sigma)
e^{-\beta H(\sigma)}}{\sum_{\sigma \in \eqm} e^{-\beta H(\sigma)}}, 
\end{equation}
and  $\mean{M}_{\text{m}}$ is similarly the mean magnetisation in the
metastable state.  The quantity $\mean{M}_{\text{m}} -
\mean{M}_{\text{eq}}$ has a physical meaning for those values of the
external field $h$ for which a metastable state exists, namely the
distance on an averaged hysteresis loop between the two branches.  

\begin{figure}
\includegraphics[scale=1]{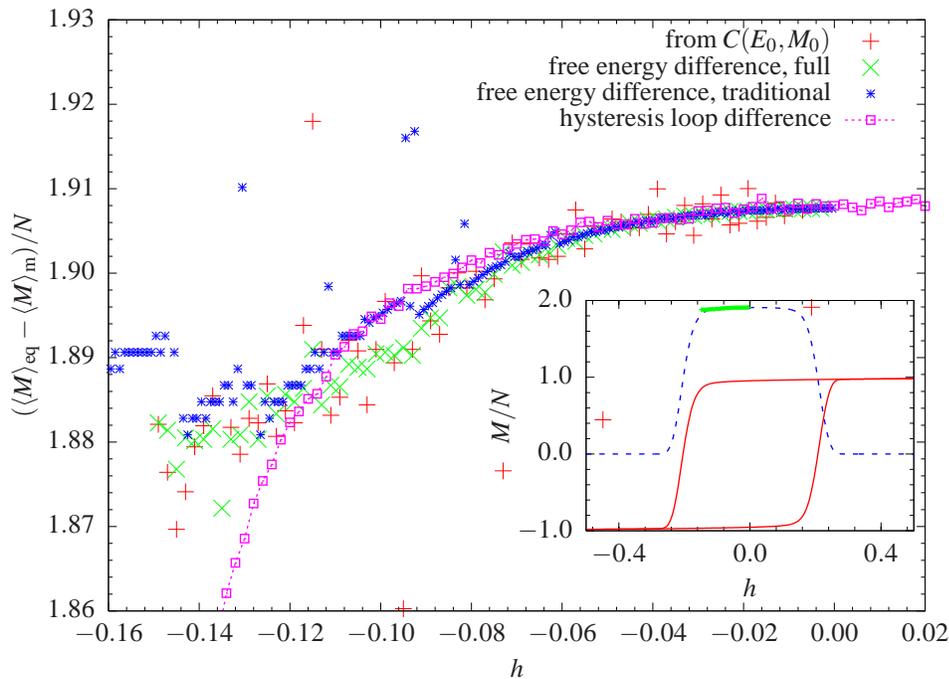}
\caption{\label{fig:hyst-comparison} Comparison of hysteresis loop
and derivative of $\ln C$ with respect to $h$, for $\beta=0.55$. For each value
of $h$, $10^7$
MCSS were used to find the metastable distribution.  The hysteresis
loop was obtained by averaging $1000$ runs, in each run increasing
$h$ in steps of $0.002$ and allowing the system to equilibrate at the
new value of $h$ for $5$ MCSS.  Shown are the data for the three
different ways of calculating $C$ referred to in the text: despite the use of
a numerical derivative, the maximum deviation of these from the hysteresis
loop data is of the order of only $2\%$. The inset
shows the complete hysteresis loop and the difference between the two
branches, as well as the data of type (i).}
\end{figure}

We evaluate $C$ as a function of $h$ in three different ways and take
the numerical derivative.  The first is $C(E_0, M_0)$ (i.e.\ $C(E,M)$
evaluated where the metastable distribution $Q_1$ attains its
maximum). The other evaluations use the free energy difference
\eqref{eq:C-as-free-energy-diff}, taking $\tmetaregion$ to be (i)
those $(E,M)$ for which $C(E,M)$ is within $\pm 1$ of $C(E_0, M_0)$,
and (ii) those $(E,M)$ for which $M$ lies on the right of the maximum
of $F(M; \beta, h)$, which is a more `traditional' method
\cite{Schulman1980}.  \figref{fig:hyst-comparison} shows $(2 / \beta)
(\partial C / \partial h)$ compared to the difference between the
heights of the two branches on a hysteresis loop.  The agreement is
reasonably good, including for the more traditional method, although
the data from $C(E_0, M_0)$ is noisy.  We remark that this provides a
possible experimental avenue for measuring a physical quantity
directly related to $C$.

\section{Several metastable states}
\label{sec:5}
In contrast to the systems we have discussed thus far, it may happen
that a given system has several metastable states: see for example
studies of a Blume--Capel model with two metastable states in
\cite{CirilloOlivieriMetastableBlumeCapel1996,
RikvoldNovotnyCompetingMetastablePRE1994}, and also
\cite{SandersUnpublished}.  In this section we extend our formalism to
describe such situations. As before, instead of focusing on a specific
example, we approach the problem through the general formalism of
Markov processes satisfying detailed balance.  Previous results in a
similar direction can be found in Refs.~\cite{GaveauSchulmanJMP1998,
GaveauSchulmanMultiplePhasesPRE2006}.

In the following, we limit ourselves to the case in which the number
of metastable states is independent of the system size $N$. Other
situations are also possible: for example, it is generally assumed
that the physics of both structural and spin glasses may be related to
the presence of a macroscopic number of metastable states
\cite{BiroliKurchanMetastableGlassy2001}. However, such a scenario
presents significant additional complexities which we do not
address. In particular, it is not clear that for such systems there
really exists an appropriate description in thermodynamic terms, as we
show in this paper for the systems we call metastable.

Since the following may well appear unnecessarily complex, let us
first explain the origin of the difficulties that may arise when
dealing with multiple metastable states. In the case in which only one
metastable state is present, there is only one eigenstate $P_{1}$,
which is essentially non-zero in the metastable region. To generalize
this to the case of $K$ isolated metastable states, all of which decay
to equilibrium, is indeed straightforward: one then finds $K$
different regions and $K$ eigenstates, one concentrated on each
region, and everything is essentially very similar to the case of a
single metastable state. The non-trivial issue arises when one
metastable state must nucleate another metastable state before it can
reach equilibrium. Under these circumstances, there is no clear
correspondence between the regions in which the $P_{\alpha}$ are
significantly different from zero and the metastable regions. We must
therefore proceed slightly differently, as follows.

We now denote by $P_{\alpha}$ all the eigenstates of the operator $L$
of the master equation (\ref{eq:2.4}) which have small relaxation
rates $\Omega_{\alpha}$. The various $\Omega_{\alpha}$ may either be
all of the same order, or differ considerably from each other. The
crucial point is that they satisfy $\Omega_{\alpha}\ll\Omega_{K+1}$,
i.e. they are all ``small'', and their number $K$ should be fixed,
independent of system size $N$.

In analogy to the case of systems with a single metastable state, we
define
\begin{eqnarray}
    C_{\alpha}(\sigma) &\defeq
    \frac{P_{\alpha}(\sigma)}{P_{0}(\sigma)}, \nonumber \\
    \left|C_{\alpha}\right| &\defeq \max_{\sigma \in
    \Gamma}\left|C_{\alpha}(\sigma) \right|,
    \label{eq:3a.2} \nonumber \\
    \Gamma_{\alpha}  &\defeq  \left\{
    \sigma:(1-\lambda_\alpha)C_{\alpha}\leq C_{\alpha}(\sigma)\leq
C_{\alpha}
    \right\}.\nonumber
\end{eqnarray}
Note that the eigenstates $P_{\alpha}(\sigma)$ are defined up to a
global sign; we choose the sign so that the $C_\alpha$ are positive.
The numbers $0<\lambda_\alpha<1$ are chosen to ensure that the sets
$\Gamma_\alpha$ are disjoint. For these states to be metastable, we
will assume that such a set of numbers exists, and that they are
$O(1)$.

Using exactly the same approach as in the previous section we can show
that $e^{\Omega_\alpha t}C_\alpha [\sigma(t)]$ is a martingale for any
$\alpha$. Defining $T_{\alpha}$ as the first time that the system
leaves $\Gamma_{\alpha}$, given that it starts with an initial
condition $\sigma_\alpha$ such that
$C_{\alpha}(\sigma_\alpha)=C_{\alpha}$, then, as before,
\begin{equation}
    \Prob(T_{\alpha}\leq t)\le\frac{1}{\lambda_\alpha} 
    \left(1-e^{-\Omega_\alpha t}\right)=O(\Omega_{\alpha}t).
    \label{eq:3a.3}
\end{equation}

Thus, if we consider the initial distribution
$\delta_{\sigma_{\alpha}}(\sigma)$, then, after an equilibration time
of order $\Omega_{K+1}^{-1}$, the system will be described by a
probability distribution given by
\begin{equation}
    Q_{\alpha}(\sigma) \defeq
P_{0}(\sigma)+\sum_{\beta=1}^{K}C_{\beta}
    (\sigma_{\alpha})P_{\beta}(\sigma).
    \label{eq:3a.5}
\end{equation}

Due to (\ref{eq:3a.3}), the probability that the process beginning at
$\sigma_{\alpha}$ leaves $\Gamma_{\alpha}$ in the relevant time range
is very small, so that
\begin{equation}
    \sum_{\sigma\notin\Gamma_{\alpha}}Q_{\alpha}(\sigma)\ll1.
    \label{eq:3a.51}
\end{equation}
Being a probability distribution, $Q_{\alpha}(\sigma)$ is
non-negative, so the above result implies that
$Q_{\alpha}(\sigma)\approx 0$ for $\sigma\notin\Gamma_{\alpha}$. Thus,
since the regions $\Gamma_\alpha$ are disjoint, we conclude that
for all $\sigma \in \Gamma$,
\begin{equation}
Q_{\alpha}(\sigma)Q_{\beta}(\sigma)\approx 0\qquad {\rm
for~~}\alpha\neq\beta
\end{equation}
and also that each $Q_{\alpha}$ is normalized over the region
$\Gamma_\alpha$. These functions play the role of $Q_{1}$ in the case
with a single metastable state.

Again, it is straightforward to show that a restricted process can be
constructed inside each $\Gamma_{\alpha}$, and that such a process
remains close to the original unrestricted process in the relevant
time range if the initial condition of both processes satisfies
$C_{\alpha}(\sigma_\alpha)=C_{\alpha}$. Thus, we can identify
\begin{equation}
Q_\alpha(\sigma)=Z_{\alpha} P_0(\sigma)\qquad(\sigma\in\Gamma_\alpha),
\label{eq:3a.510}
\end{equation}
where the constant $Z_\alpha$ is given by
\begin{equation}
Z_\alpha =
\frac{\sum_{\sigma\notin\Gamma_{\alpha}}P_0(\sigma)}{\sum_{\sigma\in
\Gamma_{\alpha}} P_0(\sigma)} \simeq
\left[
\sum_{\sigma\in
\Gamma_{\alpha}} P_0(\sigma)
\right]^{-1},
\end{equation}
in analogy to equation (\ref{eq:a.210}). We can therefore again
interpret $Z_{\alpha}$ as the partition function of the ensemble
restricted to $\Gamma_{\alpha}$.

Defined in this way, the $Q_\alpha$ are orthogonal to each other
although, being normalized as probability distributions, they are not
orthogonal to $P_0$. Nevertheless, the functions $Q_\alpha$ together
with $P_0$ still form a linearly independent basis set, and the
description of the system can be carried out in terms of these
functions, which are essentially stationary in the relevant time
range.

Now, for the slowly decaying states $Q_\alpha$ to describe
metastability, an additional condition is still required, namely
\begin{equation}
    \sum_{\sigma\in\Gamma_{\alpha}}P_{0}(\sigma)\ll 1
    \label{eq:3a.4}
\end{equation}
for all $\alpha$. Thus, the functions $Q_\alpha$ are not only assumed
to essentially be different from zero on disjoint sets, but also, the
states they describe are assumed to be extremely improbable in
equilibrium.

The aim now is to show that, if the system starts from an arbitrary
initial condition, then, with high probability, it either evolves to a
state for which the $C_{\alpha}(\sigma)$ are close to those of a
$Q_\alpha(\sigma)$ or to equilibrium. Further, this happens on a
``short'' timescale, that is, of the order of at most
$\Omega_{K+1}^{-1}$.

As before, the first step in this direction is to consider the
evolution of a system starting at an arbitrary $\sigmaz$.  After a
time of order $\Omega_{K+1}^{-1}$ has elapsed, the system will be
described by the probability distribution
\begin{equation}
    P(\sigma|\sigmaz)=P_{0}(\sigma)+
    \sum_{\alpha =1}^{K}C_{\alpha}(\sigmaz)P_{\alpha}(\sigma).
    \label{eq:3a.54}
\end{equation}
We can express the functions $P_{\alpha}(\sigma)$ in terms of the
$Q_{\alpha}(\sigma)$ and the equilibrium distribution
$P_{0}(\sigma)$, as
\begin{equation}
P_{\alpha}(\sigma)
=\sum_{\beta=1}^K
\frac{C_\alpha(\sigma_\beta)}{Z_\beta}
Q_\beta(\sigma)
-P_0(\sigma)\sum_{\beta=1}^K \frac{C_\alpha(\sigma_\beta)}{Z_\beta},
\end{equation}
where the coefficients are obtained from (\ref{eq:3a.5}) and
(\ref{eq:3a.510}) and from the orthogonality between $P_0(\sigma)$ and
$P_{\alpha}(\sigma)$, as well as from the fact that $P_0(\sigma)$ is
negligible on each $\Gamma_\alpha$.  Thus we can rewrite the
expression for $P(\sigma|\sigmaz)$ as
%\begin{widetext}
\begin{equation}
\hspace*{-60pt}
    P(\sigma|\sigmaz)=\left[1-\sum_{\beta=1}^K
    \sum_{\alpha=1}^K 
    \frac{C_\alpha
    (\sigmaz)C_\alpha(\sigma_\beta)}{Z_\beta}
    \right]P_0(\sigma)
    +\sum_{\beta=1}^K\left[\sum_{\alpha=1}^K 
    \frac{C_\alpha
    (\sigmaz)C_\alpha(\sigma_\beta)}{Z_\beta}
    \right]Q_\beta(\sigma).
\end{equation}
%\end{widetext}
The above expression means that the system has evolved to one of the
states described by a $Q_\beta$ distribution, with
probability
\begin{equation}
P(\sigma\in\Gamma_{\beta} | \sigmaz)= \sum_{\sigma\in\Gamma_{\beta}}
P(\sigma|\sigmaz) = \sum_{\alpha=1}^K \frac{C_\alpha
    (\sigmaz)C_{\alpha}(\sigma_{\beta})}{Z_{\beta}},
\label{eq:3a.533}
\end{equation}
and will decay to equilibrium with probability
\begin{equation}
P(\sigma\in\Gammaeq| \sigmaz) =1- \sum_{\beta=1}^K \sum_{\alpha=1}^K
    \frac{C_\alpha (\sigmaz)C_{\alpha}(\sigma_{\beta})}{Z_{\beta}}.
\end{equation}
Thus, for the process $\sigma(t)$
starting at $\sigma(t=0)=\sigmaz,$ the functions $C_\alpha[\sigma(t)]$
tend to the values $C_\alpha(\sigma_\beta)$ with probability
$P(\sigma\in\Gamma_{\beta} |\sigmaz)$, or to zero with probability $P
(\sigma\in\Gammaeq| \sigmaz)$, in a time of order $\Omega_{K+1}^{-1}$.

Further, note that if we choose $\sigmaz=\sigma_\gamma$ in equation
(\ref{eq:3a.533}), then we find
\begin{equation}
\sum_{\alpha=1}^K \frac{C_\alpha
(\sigma_\gamma)C_{\alpha}(\sigma_{\beta})}{Z_{\beta}}
=\delta_{\gamma\beta}.
\label{eq:3a.5335}
\end{equation} 
Otherwise, the system would quickly leave $\Gamma_{\gamma}$ even
though it had started at $\sigma_{\gamma}$, which is contrary to what
we have shown in equations (\ref{eq:3a.3}) and (\ref{eq:3a.4}).  This
can also be understood as a statement that the functions
$Z_{\alpha}^{-1/2}Q_{\alpha}$ are othonormal with respect to the
scalar product (\ref{eq:2.5}).

Thus, a system starting at an arbitrary initial state will either
decay to equilibrium or evolve to a state described by one of the
metastable distributions $Q_\alpha$. If it reaches the metastable
state $Q_\alpha$, then the function $C_\alpha(\sigma)$ for this
process will grow to $C_\alpha=C_\alpha(\sigma_\alpha)$. This occurs
in a time of order $\Omega_{K+1}^{-1}$, after which the results
pertaining to processes characterized by having the maximal value of
$C_\alpha(\sigma)$ apply. In particular, it will be very probable that
the process remains in $\Gamma_\alpha$ for a long time. Further, once
in the metastable state, it is described by the equilibrium
distribution restricted to that region of phase space, which is again
the expected physical behavior of metastable states.

Finally, note that a formula analogous to $Z_{1}=1+C^2$, which was
derived for the case of one metastable state, can be obtained as
follows. We have
\begin{equation}
Q_{\alpha}(\sigma)=P_{0}(\sigma)
\left[1+\sum_{\beta}C_{\beta}(\sigma_{\alpha})
    C_{\beta}(\sigma)\right],
    \label{eq:3.a.534}
\end{equation}
and upon substituting $\sigma_{\alpha}$ for $\sigma$ and using the
fact that
$Q_{\alpha}(\sigma_{\alpha})=Z_{\alpha}P_{0}(\sigma_{\alpha})$, one
finally obtains
\begin{equation}
    Z_{\alpha}=
1+\sum_{\beta}C_{\beta}(\sigma_{\alpha})^2.
    \label{eq:3.a.535}
\end{equation}
 
The other feature that can be discussed within this formalism is the
decay path of a metastable system. For this concept to be clear cut,
we will consider the case in which the decay rates of the metastable
states, while still small, are very different from each other. In such
a scenario, there are time scales on which the fastest metastable
state has decayed with certainty whereas no other one has. The
question then is whether we can evaluate the probability that a system
originally in the short-lived metastable state will decay to another
metastable state.

Since the eigenvalues of the evolution operator are assumed to be
ordered in increasing magnitude, we denote by $Q_K(\sigma)$ the
distribution describing the metastable state with shortest
lifetime. This distribution has support on $\Gamma_K$ which is
characterized by $C_K(\sigma)\approx C_K$. 

Now, after a time greater than $\Omega_K^{-1}$, the contributions from
$P_K(\sigma)$ vanish and the initial distribution evolves into
\begin{equation}
    Q_{K}^\prime(\sigma)=P_{0}(\sigma)+\sum_{\beta=1}^{K-1}C_{\beta}
    (\sigma_{K})P_{\beta}(\sigma),
\end{equation}
with the probability of finding the system in the initial
metastable state vanishing:
\begin{equation}
    \sum_{\sigma\in\Gamma_K} Q_{K}'(\sigma)= 0,
\end{equation}
although $Q_{K}'$ is still normalized. In particular, this means that
$Q_{K}(\sigma)=C_{K} (\sigma_{K})P_{K}(\sigma)$ for $\sigma\in \Gamma_K$. 

Expressing $Q_K'(\sigma)$ in terms of the remaining $Q_\beta(\sigma)$, we
find that the probability that the state be found in $\Gamma_\alpha$
is given by
\begin{equation}
P_{K\to \alpha}= \sum_{\beta=1}^{K-1}\frac{C_\beta(\sigma_K)
C_{\beta}(\sigma_{\alpha})}{Z_{\alpha}}
= -\frac{C_K(\sigma_K)C_{K}(\sigma_{\alpha})}
{Z_{\alpha}},
\label{p1}
\end{equation}
where we have used (\ref{eq:3a.5335}) in the last equality. 

Thus, the probability for the decay from this metastable state to
another can be obtained from the values of the coefficients
$C_\alpha(\sigma_\beta)$ appearing in the previous equation. (Note
that the above expression implies that $C_K(\sigma_\alpha)\le 0$.)
The probability that the state instead decays directly to equilibrium
is
\begin{equation}
P_{K\to {\rm equilibrium}}=1- \sum_{\alpha=1}^{K-1} P_{K\to \alpha}.
\label{p2}
\end{equation}
Thus, under certain circumstances, when the decay rates of the various
eigenmodes are very different or their corresponding metastable states
fall along different paths, the complete decay of the system can be
described as an irreversible Markov chain with transition
probabilities given by equations (\ref{p1}) and (\ref{p2}).

\section{Conclusions and Outlook}
\label{sec:6}

A fundamental issue in metastable systems concerns the possibility of 
describing them as thermodynamic equilibria (in an extended sense) of
the system at hand. The fact that they decay to an equilibrium state
which is in general quite different appears to preclude such an
approach, as do the various results concerning the impossibility of
analytically continuing the free energy beyond the coexistence curve. 
On the other hand, the use of a thermodynamic approach is routine in
the applied work on the subject. 

In this work, we have attempted to justify the thermodynamic approach 
starting from a Markovian description. While this may, at first, seem 
to be an exceedingly restrictive assumption, a moment's thought will
show the contrary: indeed, the Markovian approximation is expected to 
become reasonable on large time scales. But metastability is
essentially concerned with that time range which covers times much
larger than any microscopic relaxation time but much shorter than the 
average nucleation time (which we have, throughout the paper, called
``the relevant time range''). If such a range does not exist, or if it
is not large enough, then we may not meaningfully speak of a metastable
state. The Markovian approximation is therefore expected to be
relevant within the time range of interest. 

Further, we assume that the Markovian dynamics satisfies detailed
balance.  This last condition is not essential, and indeed, other
approaches along similar lines have been made without the assumption
of detailed balance \cite{GaveauSchulmanJMP1998,
GaveauSchulmanMultiplePhasesPRE2006}.  However, in addition to
simplifying the derivation, this assumption allows the unambiguous
identification of the probability distribution describing the
equilibrium state with that of equilibrium statistical mechanics, thus
justifying the use of concepts from equilibrium statistical mechanics
and thermodynamics in the description of metastable states.

Finally, we have limited ourselves systematically to finite systems. On
the one hand, this comes from the fact that severe technicalities
arise whenever infinite systems are considered as such. More
specifically, however, as already argued by Gaveau and Schulman
\cite{GaveauSchulmanJMP1998}, the thermodynamic limit presents some
unique difficulties for metastability: indeed, since the size of the
critical nucleating droplet remains constant as the thermodynamic
limit is attained, the time in which the first such droplet will arise
goes to zero as the system size increases to infinity.

Our definition of a metastable state includes two components: first,
it should involve an isolated eigenstate (when the system has only one
metastable state) of the master operator having an exceptionally low
eigenvalue. This eigenstate allows us to define the metastable region
in the phase space of the system. Second, we impose a technical
condition (\ref{eq:2.11}) meaning that the probability of being in a
metastable state at equilibrium is vanishingly small. The first
condition serves to discard the possibility of slow decay mediated by
long wavelength hydrodynamic modes. Indeed these usually arise as a
quasi-continuum of low-lying excitations and therefore cannot satisfy
the condition of being isolated. On the other hand, the technical
condition (\ref{eq:2.11}) is rather more difficult to understand.
Physically, however, since the properties of the metastable and
equilibrium phases are markedly different, it is clear that we should
make such a requirement of any metastable state, as was already
pointed out in \cite{PenroseLebowitz1971}.

Under the above hypotheses we have shown the following results: first,
that any initial condition whatsoever will relax, in a short time, to a
state which is either fully in the metastable region or
to equilibrium. Further, any state starting well inside the metastable
region has a very low probability of leaving it in the relevant time
range. We were also able to generalize these results to the case in
which a finite number of metastable states exist. We could not,
however, extend this to situations in which the number of metastable
states grows with the system size: this clearly cannot be done, since
it would include, among others, the case of slowly decaying
hydrodynamic modes, which correspond to a physically entirely
different situation.

A further important result allows to justify the thermodynamic
treatment: we show that if one starts inside the metastable region,
then a Markov process which reflects the system whenever it attempts
to leave the metastable region, is in fact close to the original
physical process over the relevant time range. We may therefore, for
properties which can be observed over the relevant time range, use
this ``restricted'' process instead of the original one: all the
difficulties associated with the existence of nucleation then
disappear, so we may apply the entire machinery of equilibrium
statistical mechanics to it (Green--Kubo formulae, linear response and
so on) while remaining close to the correct answer for the original
system.

The main open issue clearly concerns systems with a macroscopic number 
of low-lying eigenstates of the master operator. At least two
apparently different classes of such systems are known: on the
one hand, as we have said before, systems in which slow hydrodynamic
modes play a role. On  the other hand, both structural and spin
glasses are assumed to exhibit a large number of metastable states.
Clearly, neither can, at present, be treated by the methods presented 
here, but their extension to such systems certainly presents an
interesting challenge.

%\section*{Acknowledgements}
\ack The authors gratefully acknowledge valuable
discussions with J.L.~Lebowitz and C.~Zhou.  DPS thanks UNAM for financial
support, and the financial support of DGAPA project IN100803 is also
acknowledged.

\end{document}